\documentclass[10pt,aps,prl,groupedaddress,showpacs,twocolumn,showkeys]{revtex4-1}

\usepackage{amsmath}
\usepackage{amssymb}
\usepackage{amsfonts}
\usepackage{bm}
\usepackage{graphicx}
\usepackage{epstopdf}
\usepackage{float}
\usepackage{caption}
\usepackage{subcaption}
\usepackage{mathrsfs}
\usepackage[USenglish,american]{babel}
\usepackage{color}

\begin{document}
\title{Modal Expansion of the Scattered Field: Causality, Non-Divergence and Non-Resonant Contribution}
\author{R\'{e}mi Colom$^{1,2,*}$, Ross McPhedran$^{2}$, Brian Stout$^{1}$, Nicolas Bonod$^{1,*}$}
\affiliation{$^1$Aix Marseille Univ, CNRS, Centrale Marseille, Institut Fresnel, Marseille, France}
\affiliation{$^2$Centre for Ultrahigh-bandwidth Devices for Optical Systems (CUDOS), University of Sydney NSW 2006, Australia}
\email{remi.colom@fresnel.fr}
\email{nicolas.bonod@fresnel.fr}

\begin{abstract}
Modal analysis based on the quasi-normal modes (QNM), also called resonant states, has emerged as a promising way for modeling the resonant interaction of light with open optical cavities. However, the fields associated with QNM in open photonic cavities diverge far away from the scatterer and the possibility of expanding the scattered field with resonant contributions only has not been established. Here, we address these two issues while restricting our study to the case of a dispersionless spherical scatterer. First, we derive the rigorous pole expansion of the $T$-matrix coefficients that link the scattered to the incident fields associated with an optical resonator. This expansion evinces the existence of a non-resonant term. Second, in the time domain, the causality principle allows us to solve the problem of divergence and to derive a modal expansion of the scattered field that does not diverge far from the scatterer.
\end{abstract}

\maketitle
\section{Introduction}
Light can resonantly interact with dielectric and metallic particles that behave as 3D open cavities with radiative and possibly absorption losses. The analysis of the resonance is of uppermost importance in understanding and optimizing the process. Among the several theories developed for such an analysis, the Quasi Normal Mode (QNM) theory, also called resonant state expansion, has attracted much attention since it allows for an interpretation of the electromagnetic response of optical resonators with respect to their eigenmodes associated with complex eigenfrequencies $\omega_{\text{p},n}$ \cite{Leung1996,muljarov2011brillouin,Sauvan2013,Ge2014,powell2014resonant,Vial2014,muljarov2016exact,chen2017generalizing,Gralak2017,pick2017,mansuripur2017leaky,lalanne2018light}. Even though the use of QNM is quite recent in nanophotonics, it has a long history in quantum scattering theory, electromagnetism and nuclear physics \cite{more1971theory,more1973properties,Garcia-Calderon1997,ZelDovich1961,Kapur1938} starting from the Gamow states \cite{Gamow1928} and the singularity expansion method \cite{baum1971singularity,vincent1978singularity}. 

The imaginary part of the eigenfrequencies $\omega_{\text{p},n}$ reflects the energy decay experienced by the modes of open and passive systems. That means that under an $\exp(-i\omega t)$ time dependence, the imaginary part of $\omega_{\text{p},n}$ must be negative. This negative imaginary part of the eigenfrequencies in turn causes a divergence of the associated eigenmodes at large distances from the scatterer $r \rightarrow \infty$ as they verify outgoing boundary conditions meaning that their radial dependence asymptotically tends towards $e^{i k_{n} r}/r$, $k_{n}=\frac{\omega_{\text{p},n}}{c}$ \cite{Koenderink2010}. It is interesting to note that this problem was raised in a different context by Lamb as early as 1900 \cite{Lamb1900,beck1960physical,nussenzveig1972causality} who introduced the terminology ``exponential catastrophe'' following the pioneering work of J.J. Thomson \cite{Thomson1883,beck1960physical,nussenzveig1972causality}. Because of the divergence of these resonant states, their normalization has to be redefined \cite{ZelDovich1961,Garcia-Calderon1997,Leung1996,kristensen2015normalization,Kristensen2012,Ge2014,Sauvan2013,Doost2014,zambrana2015purcell,muljarov2016exact,Stout2017}. 
QNM analysis of the scattering problem of optical resonators has been carried out by expanding the internal field on the QNM basis in order to study the scattering problem \cite{Ge2014}. The scattered field was then computed by means of the Green function \cite{Ge2014,powell2014resonant,Powell2017}.
QNM expansions was recently used to investigate the temporal dynamics of photonic resonators \cite{faggiani2017modal,yan2018rigorous}.
However, two major problems still remain and prevent us from having a clear understanding of the QNM analysis of the scattering problem:\\
1. A fundamental problem remains regarding the divergence of the QNM fields: how is it possible to express the scattered field as a sum of QNM whereas QNM diverge when $r \rightarrow \infty$? We will show that in the time-domain, the causality principle solves the problem of divergence faced by time harmonic QNM fields. The modal expansion in the time domain of the outgoing and scattered fields reveals that a part of the scattered field keeps the same temporal dependence as the incident field (mere reflection) while the second part involves the eigenmodes of the scatterer and modifies the temporal dependence of the incident field. 

2. The other problem we want to treat is related to the expansion of the scattered field: can the scattered field of a resonator be expanded as a sum of resonant contributions only or should a non-resonant contribution also be taken into account? This question remains, although it has long been established that the internal field existing inside a spherical resonator could be expanded as a sum of QNMs only \cite{Leung1996,Lee1999a}. This result is only valid for spherically symmetric scatterers but has been assumed in most of the QNM analysis of
arbitrary shaped scatterers \cite{Ge2014,Bai2013} for which the internal field was recast as a superposition of QNMs. 
Here, we derive an expression of the scattered field possessing a non-resonant part in addition to resonant parts associated with the QNM fields. The non-resonant contribution is needed to derive convergent expansion of the scattered field. It is different from the incident field and is not required for expanding the internal field.

\section{Modal expansion of the scattering operators}
When considering light-scattering, an excitation field $\mathbf{E}_{\text{exc}}(k\mathbf{r})$ illuminates a scatterer and gives rise to a scattered field $\mathbf{E}_{\text{scat}}(k\mathbf{r})$ outside the scatterer and an internal field $\mathbf{E}_{\text{int}}(k\mathbf{r})$. 
The total field outside the scatterer is the superposition of the excitation and scattered field: $\mathbf{E}_{\text{tot}}(k\mathbf{r}) = \mathbf{E}_{\text{exc}}(k\mathbf{r}) + \mathbf{E}_{\text{scat}}(k\mathbf{r})$ and can be also decomposed as a sum of incoming $\mathbf{E}_{\text{in}}(k\mathbf{r})$ and outgoing $\mathbf{E}_{\text{out}}(k\mathbf{r})$ fields. The incoming and outgoing fields $\mathbf{E_{\text{in}}}(k\mathbf{r})$ and $\mathbf{E_{\text{out}}}(k\mathbf{r})$ are expanded on a set of incoming ($\mathbf{N}_{n,m}^{(-)}(k\mathbf{r})$, $\mathbf{M}_{n,m}^{(-)}(k\mathbf{r})$) and outgoing  ($\mathbf{N}^{(+)}(k\mathbf{r})$, $\mathbf{M}^{(+)}(k\mathbf{r})$) Vector Partial Waves (VPWs). 
Given the definitions of the incoming and outgoing VPWs (see Supplemental Material), $\mathbf{E}_{\text{in}}(k\mathbf{r})$ and $\mathbf{E}_{\text{out}}(k\mathbf{r})$ can be expressed as follows:
\begin{align}\begin{split}
\mathbf{E}_{\text{out}}(k\mathbf{r})&= E_{0}\sum_{n,m}^{\infty} s_{n,m}^{(h,+)}(\omega)\mathbf{M}_{n,m}^{(+)}(k\mathbf{r})+s_{n,m}^{(e,+)}(\omega)\mathbf{N}_{n,m}^{(+)}(k\mathbf{r})\\ 
\mathbf{E}_{\text{in}}(k\mathbf{r})& =  
E_{0}\sum_{n,m} s_{n,m}^{(h,-)}(\omega)\mathbf{M}_{n,m}^{(-)}(k\mathbf{r})+s_{n,m}^{(e,-)}(\omega)\mathbf{N}_{n,m}^{(-)}(k\mathbf{r})
\label{E_tot_exp}
\end{split}\end{align}
where $E_{0}$ is the amplitude of the field, $s_{n,m}^{(i,-)}$ (resp. $s_{n,m}^{(i,+)}$) are the coefficients of expansion of the incoming (resp. outgoing)
 field on the incoming (resp. outgoing) VPWs, with $i=(e,h)$, $e$ and $h$ denoting electric and magnetic modes, $n$ and $m$ denote the multipolar order
(see Supplemental Material). The $S$-matrix operator provides the outgoing field with respect to the incoming field: 
$\mathbf{E}_{\text{out}}(\omega)=S(\omega)\mathbf{E}_{\text{in}}(\omega)$. For spherically-symmetric scatterers, it takes the form of a diagonal matrix in a multipolar representation with coefficients defined as $S^{(i)}_{n}=s^{(i,+)}_{n,m}/s^{(i,-)}_{n,m}$. The analytic properties of the $S^{(i)}_{n}$ coefficients are linked with the energy conservation and causality. In particular, for passive media, the norm of all the 
$S$-matrix elements are bounded $\lvert S_{n}^{(e,h)}\rvert \leq 1$ (see Supplemental Material) and $\lvert S_{n}^{(e,h)}\rvert = 1$ for lossless scatterers (where $\lvert\; \rvert$ corresponds to the norm). The causality principle also has important implications on the analytic properties of the $S$-matrix 
coefficients \cite{Stout2017,Toll1956,nussenzveig1972causality}. Causality notably implies that the coefficients $S_{n}^{(e,h)}(\omega)$ have to be regular, $i.e.$ do 
not admit poles, in the upper part of the complex plane. Demonstrations of this analytic property of the $S$-matrix coefficients can be found in \cite{Toll1956,nussenzveig1972causality}. Poles satisfy the condition $\left.S_{n}^{(e,h)}
\right.^{-1}\left(\omega_{\text{p},n,\alpha}^{(e,h)}\right)=0$ and can be identified as the QNM eigenfrequencies of the scatterers.
The poles of the $S$-matrix, $\omega_{\text{p},n,\alpha}^{(e,h)}=\omega_{\text{p},n,\alpha}^{(e,h)'}+i\omega_{\text{p},n,\alpha}^{(e,h)''}$, consequently have a negative imaginary part: $\omega_{\text{p},n,\alpha}^{(e,h)''}$.  Moreover, 
energy conservation and time-reversal symmetry yield for a lossless scatterer: $S_{n}^{(e,h)}(\omega)=1/S_{n}^{(e,h)*}(\omega^{*})$ 
\cite{nussenzveig1972causality,Grigoriev2013}. As a consequence, $S_{n}^{(e,h)}(\omega)$ should also admit zeros in the upper part of the complex plane that are mirror images of  the poles of $S_{n}^{(e,h)}(\omega)$ with respect to the real axis for a lossless scatterer \cite{nussenzveig1972causality}. These zeros  $z_{n,\alpha}^{(e,h)}$ satisfy the condition $S_{n}^{(e,h)}(z_{n,\alpha}^{(e,h)})=0$ and are associated with modes verifying incoming boundary conditions \cite{Grigoriev2013,Baranov2017}. The reality of the field and the time reversal symmetry impose that the mirror images of $\omega_{\text{p},n,\alpha}$ and $z_{n,\alpha}$ with respect of the imaginary axis: $-\omega_{\text{p},n,\alpha}^{(e,h)*}$ and $-z_{n,\alpha}^{(e,h)*}$ are also poles and zeros of $S_{n}^{(e,h)}$. The S-matrix coefficients can be cast as a product between a holomorphic function $S_{\text{R},n}^{(e,h)}(\omega)$ and an exponential phase factor $e^{-2i\frac{\omega}{c}R}$: $S_{n}^{(e,h)}(\omega)=e^{-2i\frac{\omega}{c}R}S_{\text{R},n}^{(e,h)}(\omega)$. The prefactor $\exp (-2ikR)$ expresses the causality requirement, and arises from the ratio of outgoing and incoming zero-order Hankel functions at the sphere surface. This term will be necessary to guarantee the convergence of the sum form in Eq.(\ref{S_n_express}) when $M\rightarrow\infty$. These considerations lead to the following infinite-product form of the S-matrix coefficients \cite{Stout2017,Grigoriev2013,VanKampen1953,Toll1956,nussenzveig1972causality}(see also Supplemental Material):

\begin{align}\begin{split}
S^{(e)}_{n}(\omega)&=(-1)^{n+1}e^{-2ikR}\prod_{\alpha=-\infty}^{\infty}\frac{\omega-z_{n,\alpha}^{(e)}}{\omega-\omega_{\text{p},n,\alpha}^{(e)}}\\
S^{(h)}_{n}(\omega)&=(-1)^{n}e^{-2ikR}\prod_{\alpha=-\infty}^{\infty}\frac{\omega-z_{n,\alpha}^{(h)}}{\omega-\omega_{\text{p},n,\alpha}^{(h)}}
\label{Weier_S}
\end{split}
\end{align}

where $k=\frac{\omega}{c}$. This infinite product converges because of the link between the positions of zeros and poles. The poles and zeros with a positive index $\alpha$ have a positive real part while those with a negative index are their mirror images with respect to the real axis (a pole on imaginary axis can be designated $\alpha=0$). If we expand $S_{\text{R},n}^{(e,h)}(\omega)$ into partial fractions, it is possible to transform the product in Eq. (\ref{Weier_S}) into an infinite sum \cite{VanKampen1953,Grigoriev2013}. In practice, one has to truncate this sum to a finite number $M$ of terms. This leads to the following expression of $S^{(e,h)}_{n}(\omega)$ (see Supplemental Material):

\begin{equation}
S_{n}^{(i)}(\omega) \simeq e^{-2ikR}\left(S_{\text{nr},n}^{(i)}+ \sum_{\alpha = -M}^{M}\frac{r_{n,\alpha}^{(i)}}{\omega-\omega_{\text{p},n,\alpha}^{(i)}}\right)
\label{S_n_express}
\end{equation}
where $i=(e,h)$ and $r_{n,\alpha}^{(i)}$ is the residue of $S_{\text{R},n}^{(i)}$ at the poles $\omega_{\text{p},n,\alpha}^{(i)}$ and the non-resonant term $S_{\text{nr},n}^{(i)}=1+\sum_{\alpha= -M}^{M}\frac{r_{n,\alpha}^{(i)}}{\omega_{\text{p},n,\alpha}^{(i)}}$. \\

\begin{figure}[h!]
\begin{center}
\includegraphics[width=0.5\textwidth]{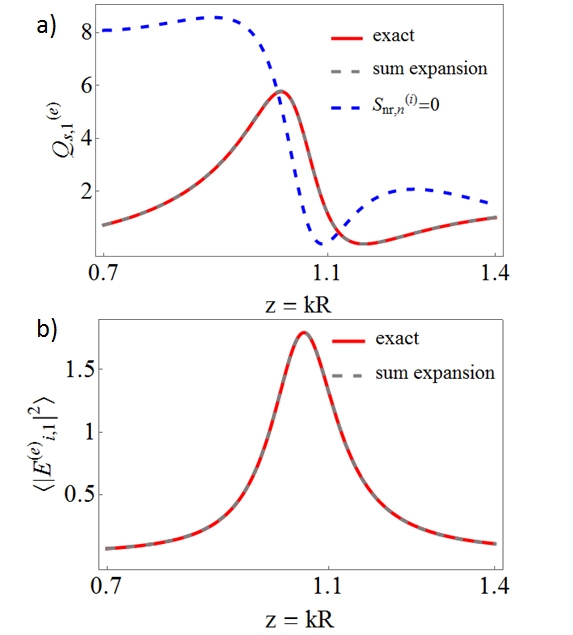}
\end{center}

\caption{(a) Dipolar electric partial scattering efficiency $Q_{1}^{(e)}=\frac{6}{z^{2}}\lvert T_{1}^{(e)}\rvert^{2}$ and (b) averaged square modulus of the internal electric field with  $\lvert \mathbf{E}_0 \rvert=1$ for an $\varepsilon = 16$ spherical scatterer in air: exact calculations (red solid curve), Eq. (\ref{T_n_express}) with $M=100$ poles for (a) and Eq. (\ref{part_fract_Xhi_Omeg}) with $M=10$ poles for (b) (gray dashed line), Eq. (\ref{T_n_express}) with $M=100$ for (a) while assuming $S_{\text{nr},n}^{(e,h)} = 0$ (blue dashed curve).}
\label{Fig:Scat_eff}
\end{figure}

So far, the use of the S-matrix formalism has proved very useful for deducing the analytic properties of the scattering matrix coefficients from general
properties like causality and energy conservation. However, the T-matrix formalism will be preferred in the following as it provides a more intuive
description of the scattering problem by linking the excitation field to the scattered field. The excitation field 
$\mathbf{E_{\text{exc}}}$ will be expanded on the set of regular VPWs 
($\mathbf{N}_{n,m}^{(1)}(k\mathbf{r})$,$\mathbf{M}_{n,m}^{(1)}(k\mathbf{r})$) with the corresponding coefficients of expansion ($e_{n,m}^{(e)}(\omega)$,$e_{n,m}^{(h)}(\omega)$), while coefficients of expansion of the scattered field $\mathbf{E}_{\text{scat}}$ on the outgoing VPWs ($\mathbf{N}^{(+)}(k\mathbf{r})$,$\mathbf{M}^{(+)}(k\mathbf{r})$) will be denoted ($f_{n,m}^{(e)}(\omega)$,$f_{n,m}^{(h)}(\omega)$) (see Supplemental Material). The T-matrix is also diagonal for a spherically symmetric scatterer and its coefficients are defined as follows: $T_{n}^{(e,h)}(\omega) = f_{n,m}^{(e,h)}(\omega)/e_{n,m}^{(e,h)}(\omega)$. A pole expansion of the T-matrix coefficient can also be derived from Eq. (\ref{S_n_express}) by means of the following relation $T = \frac{S-I}{2}$ \cite{Colom2016}, where $I$ is the identity matrix, leading to:
\begin{equation}
T^{(i)}_{n}(\omega) \simeq A_n^{(i)}(\omega)+B_n^{(i)}(\omega)\sum_{\alpha = -M}^{M}\frac{r_{n,\alpha}^{(i)}}{\omega-\omega_{\text{p},n,\alpha}^{(i)}}\\ 
\label{T_n_express}
\end{equation}

where $i=(e,h)$, $A_n^{(i)}(\omega)=\frac{e^{-2ikR}S_{\text{nr},n}^{(i)}-1}{2}$ and $B_n^{(i)}(\omega)=\frac{e^{-2ikR}}{2}$. 
This pole expansion of the T-matrix coefficients allows to determine the spectral response of a scatterer over a broad range of frequencies from a discrete set of QNMs. This is illustrated in Fig. \ref{Fig:Scat_eff}a where the accuracy of this pole expansion and the importance of these non-resonant terms are assessed. We compare in particular the dipolar electric partial scattering efficiency: $Q_{1}^{(e)}=\frac{6}{z^{2}}\lvert T_{1}^{(e)}\rvert^{2}$ obtained by exact calculation on one side and using Eq. (\ref{T_n_express}) with $M=100$ for a spherical scatterer of dielectric permittivity $\varepsilon = 16$ in air as it is close to the permittivity of silicon in the visible spectrum \cite{zambrana2015purcell}. Poles were found by using their asympotic values \cite{Stout2017} along with pole finding methods ($\varepsilon = 16$ in the whole complex $\omega$-plane). A good agreement between the exact calculations and Eq. (\ref{T_n_express}) is found. Fig. \ref{Fig:Scat_eff}a) also illustrates the importance of the non-resonant term in Eq. (\ref{T_n_express}) since a poor prediction of the electric dipolar partial scattering efficiency is obtained when neglecting it even when a large number of poles is taken into account. Similar results can be obtained for other multipolar orders. The link between the pole expansion of the T-matrix coefficients and the QNM expansion of the scattered field will be clarified in the second part of this study.

Let us now extend this work to the internal field, the field inside the resonator, by introducing the $\Xi$ matrix that relates the internal field to the incoming field. Let ($u_{n,m}^{(e)}(\omega)$,$u_{n,m}^{(h)}(\omega)$) be the coefficients of the internal field on the regular VPWs ($\mathbf{N}_{n,m}^{(1)}(k_{s}\mathbf{r})$,$\mathbf{M}_{n,m}^{(1)}(k_{s}\mathbf{r})$) where $k_{s}=\sqrt{\varepsilon_{s}}\frac{\omega}{c}$. The coefficients are then simply defined as: $\Xi_{n}^{(i)}(\omega)=\frac{u_{n,m}^{(i)}(\omega)}{s_{n,m}^{(i,-)}(\omega)}$. Similarly, it is useful to introduce the $\Omega$ matrix that relates the excitation field to the internal field and whose coefficients are defined as: $\Omega_{n}^{(i)}(\omega)=\frac{u_{n,m}^{(i)}(\omega)}{e_{n,m}^{(i)}(\omega)}$, with $\Omega_{n}^{(i)}(\omega)=\frac{\Xi_{n}^{(i)}(\omega)}{2}$. A pole expansion can also be derived for  $\Xi_{n}^{(i)}(\omega)$ and $\Omega_{n}^{(i)}(\omega)$. Energy conservation does not require the coefficients $\Xi_{n}^{(i)}(\omega)$ to be unitary. Consequently, and in contrast to $S_{n}^{(i)}(\omega)$, these coefficients do not possess zeros that are symmetric to their poles with respect to the real axis. One can show that $\Xi_{n}^{(i)}(\omega)$ and $\Omega_{n}^{(i)}(\omega)$ admit the following pole expansion (see Supplemental Material):

\begin{equation}
\Xi_{n}^{(i)}(\omega) =  \sum_{\alpha=-\infty}^{\infty} \frac{r_{\Xi,n,\alpha}^{(i)}}{\omega-\omega_{\text{p},n,\alpha}^{(i)}} \equiv 2 
\, \Omega_{n}^{(e,h)}(\omega)
\label{part_fract_Xhi_Omeg} 
\end{equation}
where $i = e$ or $h$. Unlike the pole expansions of $S_{n}^{(e,h)}(\omega)$ and $T^{(e,h)}_{n}(\omega)$, there is no non-resonant term in the pole expansion of $\Xi_{n}^{(e,h)}(\omega)$ and $\Omega_{n}^{(e,h)}(\omega)$.
To study the accuracy of this pole expansion, we calculate the averaged square internal field of a spherical resonator excited by a plane wave \cite{Stout2017} (see Supplemental Material): $\frac{\langle \lvert \mathbf{E}_{int} \rvert^{2}\rangle}{\lvert\mathbf{E}_{0} \rvert^{2}}=\sum_{n=1}^{\infty}\frac{\langle \lvert \mathbf{E}_{i,n}^{(e)} \rvert^{2}\rangle}{\lvert\mathbf{E}_{0} \rvert^{2}}+\frac{\langle \lvert \mathbf{E}_{i,n}^{(h)} \rvert^{2}\rangle}{\lvert\mathbf{E}_{0} \rvert^{2}}$. This expression is plotted in Fig. \ref{Fig:Scat_eff}b with $\varepsilon_{s} = 16$ while considering the electric dipole contribution only, $i.e. \frac{\langle \lvert \mathbf{E}_{i,1}^{(e)} \rvert^{2}\rangle}{\lvert\mathbf{E}_{0} \rvert^{2}}$.

Fig. \ref{Fig:Scat_eff}b shows a good agreement between exact calculations and the predictions obtained while using Eq. \ref{part_fract_Xhi_Omeg} with 20 poles (10 poles with a positive real part along with their symmetric with respect to the imaginary axis).\\

\section{Causality and time domain}
Let us now derive the expression of the outgoing and scattered fields in terms of the QNM fields. We will in particular show how the divergence of these QNM fields can be dealt with. That is why we will here focus on the far-field region where this divergence occurs. The modes of spherically symmetric scatterers have to belong to one type of VPWs. Let us denote the electric-type (resp. magnetic-type) modes of the $n-th$ multipolar order $E_{n,m,\alpha}^{(e)}\left(\mathbf{r}\right)$ (resp. $E_{n,m,\alpha}^{(h)}\left(\mathbf{r}\right)$). Outside the scatterer, $i.e.$ for $r > R$: $\mathbf{E}_{n,m,\alpha}^{(e)}\left(\mathbf{r}\right) \propto \mathbf{N}_{n,m}^{(+)}\left(\frac{\omega_{\text{p},n,\alpha}^{(e)}}{c}\mathbf{r}\right)$ and  $\mathbf{E}_{n,m,\alpha}^{(h)}\left(\mathbf{r}\right) \propto \mathbf{M}_{n,m}^{(+)}\left(\frac{\omega_{\text{p},n,\alpha}^{(h)}}{c}\mathbf{r}\right)$. Moreover, in the far-field region ($i.e.$ when $r\rightarrow\infty$), they possess the following asymptotic expressions $\mathbf{E}_{n,m,\alpha}^{(e),\text{FF}}\left(\mathbf{r}\right)$ and $\mathbf{E}_{n,m,\alpha}^{(h),\text{FF}}\left(\mathbf{r}\right)$ (see Supplementary Material): 
\begin{align}\begin{split}
\mathbf{E}_{n,m,\alpha}^{(e),\text{FF}}\left(\mathbf{r}\right) &\propto e^{i\frac{\omega_{\text{p},n,\alpha}^{(e)}}{c}r}\mathbf{Z}_{n,m}(\theta,\phi)\\
\mathbf{E}_{n,m,\alpha}^{(h),\text{FF}}\left(\mathbf{r}\right) &\propto e^{i\frac{\omega_{\text{p},n,\alpha}^{(h)}}{c}r}\mathbf{X}_{n,m}(\theta,\phi) \;.
\end{split}\end{align}

where $\mathbf{X}_{n,m}(\theta,\phi)$ and $\mathbf{Z}_{n,m}(\theta,\phi)$ are vector spherical harmonics (see Appendix A of \cite{Stout11}). Since $e^{i\frac{\omega_{\text{p},n,\alpha}^{(i)}}{c}r}=e^{i \frac{\omega_{\text{p},n,\alpha}^{(i)'}r}{c}}e^{-\frac{\omega_{\text{p},n,\alpha}^{(i)''}r}{c}}$ and as, due to causality, $\omega_{\text{p},n,\alpha}^{(i)''}< 0$, $E_{n,m,\alpha}^{(i)}\left(\mathbf{r}\right)$ is exponentially diverging as $r\rightarrow\infty$ ($i=e,h$). 
One may believe that this divergence could hinder the derivation of an expression of the scattered field in terms of ($E_{n,m,\alpha}^{(e)}$,$E_{n,m,\alpha}^{(h)}$). However, this divergence can be understood by noting that, due to causality, considering a scattered field at $r\rightarrow\infty$ amounts to assuming that the system was excited at $t\rightarrow - \infty$ \cite{nussenzveig1972causality,Lamb1900}. 

In what follows, we will show that a divergence free expansion of the scattered and outgoing fields of 3D open optical cavities in the time domain can be obtained. It will require the use of a causal incoming and excitation fields, $i.e.$ fields with a sharp cut-off in the time domain \cite{nussenzveig1972causality}, together with rigorous calculations based on the theorem of residues. Causality can be better understood in the framework of the $S$ matrix while considering incoming and outgoing waves in the far-field region. The incoming field defined in Eq. (\ref{E_tot_exp}) takes the following asymptotic expression in the far-field region that is denoted $\mathbf{E}_{\text{in}}^{\text{FF}}$:
\begin{eqnarray}
\begin{aligned}
&\mathbf{E}_{\text{in}}^{\text{FF}}\left(\mathbf{r},\omega\right) = E_{0}\sum_{n=1}^{\infty}\sum_{m=-n}^{n}\left(s_{n,m}^{(h,-)}(\omega)i^{n+1}\frac{e^{-ikr}}{kr}\mathbf{X}_{n,m}(\theta,\phi)\right. \\
& \left. - s_{n,m}^{(e,-)}(\omega) i^{n+2} \frac{e^{-ikr}}{kr}\mathbf{Z}_{n,m}(\theta,\phi)\right)
\end{aligned}
\label{in_electromag_field_fd}
\end{eqnarray}
where $s_{n,m}^{(i,-)}(\omega)=s_{n,m}^{(i,-)}g(\omega)$, $s_{n,m}^{(i,-)}$ being the coefficient of expansion of the incoming field on the VPWs and $g(\omega)$ being the Fourier transform of the temporal dependence of the incoming field $g(t)$.
Let us take the Fourier transform of one of the terms in this sum:
\begin{eqnarray}
\begin {aligned}
&\mathbf{E}_{\text{in},n,m}^{(e),\text{FF}}(\mathbf{r},t) = -\frac{i^{n+2}E_{0}}{2\pi} \int_{-\infty}^{+\infty} s_{n,m}^{(e,-)}(\omega) \frac{e^{-ikr}}{kr} \times\\
&\mathbf{Z}_{n,m}(\theta,\phi) e^{-i\omega t}d\omega \;.
\end{aligned}
\end{eqnarray}
As shown in the Supplemental Material, we choose the time dependence $g(t) = H(t)e^{-i\omega_0 t}$ for the incoming field since it corresponds to an incoming causal wavefront. This causal wavefront can be built up by choosing the following frequency dependence for $s_{n,m}^{(e,-)}(\omega)$: $s_{n,m}^{(e,-)}(\omega) = \lim_{\eta\rightarrow 0^{+}} s_{n,m}^{(e,-)}\frac{\omega}{\omega_{0}}\frac{i e^{i \omega t_{s}}}{\omega - \omega_{0}+i\eta}$ (with $s_{n,m}^{(e,-)}$ a constant) as it is the Fourier transform of $g(t)$. 
In a sake of generality, an additional phase factor $e^{i \omega t_{s}}$ is introduced to yield an additional time delay $t_s$. This definition of $s_{n,m}^{(e,-)}(\omega)$ can be used in the expression of $\mathbf{E}_{\text{in},n,m}^{(e),\text{FF}}(\mathbf{r},t)$ (see Supplemental Material): 
\begin{eqnarray}
\begin{aligned}
&\mathbf{E}_{\text{in},n,m}^{(e),\text{FF}}(\mathbf{r}_{0},t) = -H\left(t - t_{s}+\frac{r_{0}}{c}\right)i^{n+2}E_{0}s_{n,m}^{(e,-)}\times\\
&\frac{e^{-i k_{0} r_{0}}}{k_{0} r_{0}}\mathbf{Z}_{n,m}(\theta,\phi) e^{-i \omega_{0}\left(t - t_{s}\right)}
\end{aligned} \;,
\label{in_n_m_td_5}
\end{eqnarray}
where $H$ is the Heaviside step distribution. The time-dependent outgoing field in the far-field region can be calculated with respect to the $S$-matrix coefficients of the incoming field: $\mathbf{E}_{\text{out},n,m}^{(e),\text{FF}}(\mathbf{r},t) = \frac{(-i)^{n+2}E_{0}}{2\pi} \int_{-\infty}^{+\infty}  S_{n}^{(e)}(\omega) s_{n,m}^{(e,-)}(\omega) \frac{e^{i k r}}{k r}\mathbf{Z}_{n,m}(\theta,\phi) e^{-i\omega t}d\omega$.

\begin{figure}[H]
\begin{center}
\includegraphics[width=0.5\textwidth]{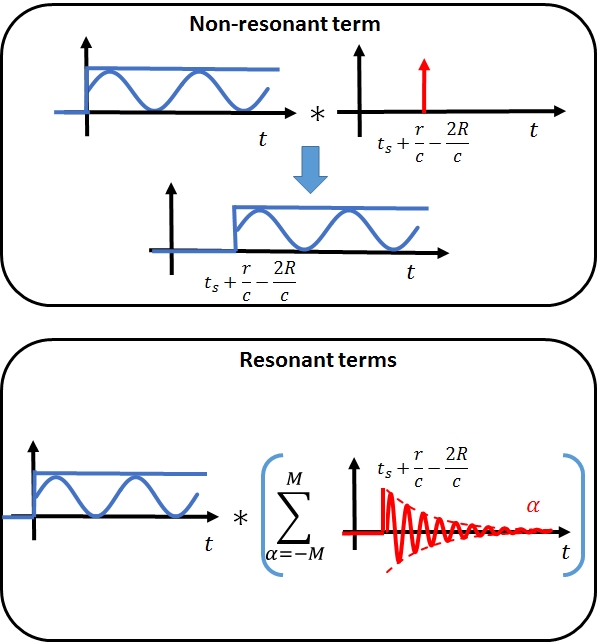}
\end{center}
\caption{Schematic representation of non-resonant and resonant terms in Eq.\ref{E_out_td} }
\label{graph}
\end{figure}

The easiest way to calculate $\mathbf{E}_{\text{out},n,m}^{(e),\text{FF}}(\mathbf{r},t)$ is to use the convolution theorem associated with the theorem of residues:

\begin{widetext}

\begin{eqnarray}
\begin{aligned}
\label{E_out_td}
\mathbf{E}_{\text{out},n,m}^{(e),\text{FF}}\left(\mathbf{r},t\right) = \frac{\mathbf{C}_{n,m}(\theta,\phi)}{k_{0}r} g(t)\ast\left(S_{\text{nr},n}^{(e)}\delta\left(t-\tau \right)- iH\left(t-\tau\right)\sum_{\alpha = -M}^{M}r_{n,\alpha}^{(e)}e^{-i \frac{\omega_{\text{p},n,\alpha}^{(e)}}{c} \left(t-\tau\right)}\right)
\end{aligned}
\end{eqnarray}
\end{widetext}

with $\mathbf{C}_{n,m}(\theta,\phi) = (-i)^{n+2}E_{0}s_{n,m}^{(e,-)}\mathbf{Z}_{n,m}(\theta,\phi)$; $g(t) = H(t)e^{-i\omega_0 t}$, and $\tau=t_{s}+\frac{r}{c}-\frac{2R}{c}$. Two contributions can be identified in Eq. (\ref{E_out_td}): ($i$) The non-resonant contribution equal to the convolution between the temporal dependence of the incoming field $g(t)$ and the Dirac distribution $\delta\left(t-t_{s}-\frac{r}{c}+\frac{2R}{c}\right)$. This contribution corresponds to a mere reflection of the incoming field by the surface of the scatterer. Importantly, it keeps the same temporal dependence as the incoming field which means that it does not distort the incoming field that is simply translated in time by a factor $\tau$; ($ii$) The resonant contributions that correspond to the convolution between $g(t)$ and a sum of exponentially decreasing terms that are characteristic of the response of each mode. This resonant response strongly distorts the temporal dependence of the incoming field which was expected since it results from a resonant interaction between light and eigen-modes of the cavity. These two contributions are sketched in Fig.\ref{graph}. 

The scattered field can finally be obtained by removing the outgoing part of the excitation field from $\mathbf{E}_{\text{out},n,m}^{(e),\text{FF}}\left(\mathbf{r},t\right)$ in Eq. (\ref{E_out_td}):
\begin{eqnarray}
\begin{aligned}
&\mathbf{E}_{\text{scat},n,m}^{(e),\text{FF}}(\mathbf{r},t) = \frac{\mathbf{D}_{n,m}(\theta,\phi)}{2k_{0}r}  g(t) \ast \\
&\left(S_{\text{nr},n}^{(e)}\delta\left(t-\tau\right)-\delta\left(t-t_{s}-\frac{r}{c}\right)\right.\\ 
&\left.- iH\left(t-\tau\right)\sum_{\alpha = -M}^{M}r_{n,\alpha}^{(e)}e^{-i \omega_{\text{p},n,\alpha}^{(e)} \left(t-\tau\right)}\right)
\end{aligned}
\label{scat_n_m_td_2}
\end{eqnarray}
with $\mathbf{D}_{n,m}(\theta,\phi)=(-i)^{n+2}E_{0}e_{n,m}^{(e)}\mathbf{Z}_{n,m}(\theta,\phi)$.

The scattered field in the time domain results from a convolution between the excitation field $g(t)$ and the function characterizing the response of the scatterer, described by two terms. The first term, $S_{\text{nr},n}^{(e)}\delta\left(t-\tau\right)-\delta\left(t-t_{s}-\frac{r}{c}\right)$, is associated with the non-resonant part $S_{\text{nr},n}^{(e)}$ of the $S$-matrix coefficients from which is substracted the outgoing part of the excitation field. This evinces the fact that the non-resonant part does not include the excitation field. The second term includes the far-field limit of the QNM fields $\mathbf{E}_{n,m,\alpha}^{(e),\text{FF}}\left(\mathbf{r}\right) \propto e^{i\frac{\omega_{\text{p},n,\alpha}^{(e)}}{c}r}\mathbf{Z}_{n,m}(\theta,\phi)$. The Heaviside step function in front of the sum prevents the field from diverging when $r \rightarrow \infty$. The Heaviside distribution results from causality and simply means that, given the initial conditions imposed on the incoming field in Eq. (\ref{in_n_m_td_5}), the outgoing field has only been able to propagate up to a distance $r = c\left(t-t_{s}\right) + 2R$. Due to the Heaviside distribution, the outgoing field is different from zero only when $t-t_{s}-\frac{r}
{c}+\frac{2R}{c}  \geq 0$. Therefore the term $e^{\omega_{\text{p},n,\alpha}^{(e)''} \left(t-t_{s}-\frac{r}{c}+\frac{2R}{c}\right)}<1$ since one has both $\omega_{\text{p},n,\alpha}^{(i)''}< 0$ and $t-t_{s}-\frac{r}{c}+\frac{2R}{c}  \geq 0$. and consequently this term does not diverge.

\section{Conclusion}
To conclude, in addition to the resonant contributions due to the excitation of QNM of the scatterer by the excitation field, a non-resonant contribution must be taken into account in the expansion of the scattered field. This result has been demonstrated by deriving pole expansions of the S and T matrix coefficients of a dispersionless 3D scatterer. We benefited from the symmetry of the spherical scatterers to provide the explicit expressions of both resonant and non-resonant contributions of the scattered field expansion. Then, the expansion of the scattered field in the time domain was computed by means of an inverse Fourier transform of the frequency-dependent scattered field obtained from the pole expansion of the T-matrix coefficients. Rigorous calculations based on the theorem of residues together with causality principle lead to QNM expansion terms that do not diverge far from the scatterer. This result obtained in the framework of the multipolar theory can, on physical grounds, be generalized to arbitrarily shaped scatterers. Moreover, QNM theory is applicable to an extremely broad range of physical studies far beyond optics.

\section{acknowledgments}
Research conducted within the context of the International Associated Laboratory ALPhFA: Associated Laboratory for Photonics between France and Australia. This work has been carried out thanks to the support of the A*MIDEX project (no. ANR-11-IDEX-0001-02) funded by the Investissements d'Avenir French Government program, managed by the French National Research Agency (ANR). The authors acknowledge Emmanuel Lassalle for fruitful discussions.

\end{document}